\documentstyle[graphicx,preprint,prb,aps]{revtex}
\tightenlines
\begin{document}
\draft
\title{
Vertically coupled quantum dots in the local
spin-density functional theory
}
\author{Mart\'{\i} Pi, Agust\'{\i} Emperador, and
Manuel Barranco}
\address{Departament d'Estructura i Constituents de la Mat\`eria,
Facultat de F\'{\i}sica, \\
Universitat de Barcelona, E-08028 Barcelona, Spain}
\author{Francesca Garcias}
\address{Departament de F\'{\i}sica, Facultat de Ci\`encies,\\
Universitat de les Illes Balears, E-07071 Palma de Mallorca, Spain}
\date{\today}

\maketitle

\begin{abstract}

We have investigated the
structure  of double quantum dots vertically coupled at zero magnetic
field within local spin-density functional theory.
The dots are identical and have a
finite width, and the whole system is axially symmetric. We first
discuss the effect of thickness on the addition spectrum  of
one single dot. Next, we describe the structure of coupled dots as a
function of the interdot distance for different electron numbers.
Addition spectra, Hund's rule and molecular-type configurations are
discussed.  It is shown that self-interaction corrections to the
density functional results  do not play a very important role in
the calculated addition spectra.

\end{abstract}

%
% 71.15.Mb: Density functional theory, local density approximation.
% 85.30.Vw: Low-dimensional quantum devices (quantum dots,
%           quantum wires, etc...)
% 73.20.Dx: Electron states in low- dimensional structures
%           (superlattices, quantum well structures and multilayers)
% 36.40.Ei: Low-dimensional quantum devices (quantum dots,
%           quantum wires, etc...)
%
%\narrowtext

\section{Introduction}

The study of systems with a small number ($N$) of electrons
confined to a quasi two-dimensional (2D) semiconductor
quantum dot (QD) constitutes a subject of growing
interest (see for instance Refs. \onlinecite{Jac98,And98}
and references therein).
One reason for this interest is that their
electronic properties can be selected with some freedom
by tailoring the shape of the lateral confining potential.
In this sense, they are  often referred to as artificial atoms.
Recently, circular and elliptically disk-shaped QD's have been built
in a very clean way, and their properties have been thoroughly
studied
\cite{Tar96,Aus98,Tar98,Sas98,Oos99,Aus99,Rei99}
as a function of $N$.

Usually, QD's are described as  true 2D systems.
This seems  justified, as it has
been widely tested by comparing theory with experiment.
However, this comparison may be obscured by the fact that
the parameters defining the confining
potential are adjusted and might mask 3D effects.
However, complex 2D calculations have been carried out without further
restrictions or imposed symmetries.
\cite{Aus99,Rei99,Kos97,Yan99,Pue99} Full 3D calculations
also exist in the Hartree or Hartree-Fock (HF) approximations
for very few electrons (see for instance Refs.
\onlinecite{Kum90,Nat99}), and also within the density functional
theory.\cite{Sto96,Nag97,Lee98,Lee99}

A systematic study of the  role of
dimensionality in QD structure has been presented
by Rontani et al.\cite{Ron98,Ron99} In spite of the success
of 2D models in describing the properties of QD's subjected
to perpendicular  magnetic fields, the width in
the growth direction of most experimentally studied QD's is around
one order of magnitude smaller than their typical radius.
The effect of  a small but finite vertical extension on the QD
structure is worth studying considering that a recent
extension of single QD studies to vertically coupled
quantum dots\cite{Sch97} may render including
the $z$ extension of the constituent dots  unavoidable
to describe the experimental data.\cite{Aus98,Tar98}

Vertically coupled dots, also called artificial or quantum dot
molecules, have been theoretically addressed in a number of works.
\cite{Ron99a,Pal95,May97,Par97,Par98,Par00} Only in
Ref. \onlinecite{Ron99a} has the $z$ extension of the constituent QD's
been taken into account; in the other references, it was
neglected and consequently, their results cannot be reliable when
the interdot distance is comparable to their $z$ extension,
which is an interesting physical situation.\cite{Aus98,Tar98}

In this work we present a study of vertically coupled, cylindrical
QD's in the local spin-density functional theory (LSDFT).
We restrict our description  to axially symmetric configurations
and identical QD's, and limit ourselves to zero magnetic
field $B$.
While these two latter conditions are easy to relax without
increasing the amount of numerical work too much,  axial
symmetry breaking would require a more demanding 3D calculation.

Our scheme is based on the application of the so-called\cite{Dav80}
imaginary-time method (ITM). It requires a discretization
of the Kohn-Sham (KS) equations on a spatial mesh that can be
straightforwardly implemented on a personal computer, and avoids
expansion of the single-particle (sp) wave functions in a basis,
 large matrix diagonalizations and
tests of the stability of the results against changes in the size of
the basis.

In contrast to standard methods for computing molecular structure
\cite{Wei78}, we do
not postulate from the start that the artificial molecular
orbitals can be expressed as linear combinations of artificial
atomic orbitals (LCAAO) and consequently, we  avoid making
 approximations such as  complete neglect
of differential overlap (CNDO)  where the individual wave
functions of electrons associated with different artificial atoms
are taken as being orthonormal, as happens when the QD's are either
two dimensional  or far apart.

The present approach to the description of vertically coupled QD's
constitutes a clear improvement on
previous LSDFT calculations\cite{Par00} in which the electrons are
located on one of the dots, and as a consequence,
can only be electrostatically coupled, even when they lie
at short distances. With respect to
the generalized Hubbard model plus the diagonalization scheme of
Ref. \onlinecite{Ron99a}, the present improvement is the possible
 application of LSDFT to systems with large numbers of electrons.
Moreover, one could think of a
trivial generalization to non-identical dots, paying the
well known token that LSDFT  treats the exchange Coulomb term
locally, and that LSDFT configurations are usually mixtures of
many-electron states with the same value
of the total spin projection $S_z$, however, with different total spin
$S$. For small systems, this drawback (which is also inherent
to the generalized Hubbard approach\cite{Ron99a}) can be removed
with a shell-model calculation in a restricted
sp valence space.\cite{Bar97,Hir99} Another difference between
 our approach and the Hubbard model is that even if the
exchange energy is treated locally, it is not restricted to involving
only electrons located on the same dot. This might be
advantageous in the strong coupling case when the dots are close together
 and the quantum mechanical coupling due to electron
exchange is important.

This paper is organized as follows. In Section II we present the
formalism and the essentials of the ITM. Results for one single
thick QD and for two coupled dots are presented in Section III,
and a summary is presented in Section IV.
In an Appendix we present the self-interaction correction\cite{Per81}
to the density functional results obtained for two extreme interdot
distances and several electron numbers.

\section{Methodological approach}

Within LSDFT, the ground-state  of the system is
obtained by solving the Kohn-Sham equations.
The problem is simplified by the imposed axial symmetry
around the $z$ axis, which allows one to write the
sp wave functions as
$\phi_{nl\sigma}(r,z,\theta,\sigma)=
u_{nl\sigma}(r,z) e^{-\imath l \theta} \chi_{\sigma}$ with
$l =0, \pm 1, \pm 2 \ldots$, where $-l$ is the projection of the
sp orbital angular momentum on the symmetry axis.

We have used effective atomic units
$\hbar=e^2/\epsilon=m=$1, where
$\epsilon$ is the dielectric constant,
and $m$  the electron effective mass. In units of the bare
electron  mass $m_e$ one has $ m = m^* m_e$.
In this system, the length unit is the effective
Bohr radius $a_0^* = a_0\epsilon/m^*$,
and the energy unit is the effective Hartree $H^* = H  m^*/\epsilon^2$.
In the numerical applications we have considered
GaAs, for which  we have taken $\epsilon$ = 12.4, and  $m^*$ = 0.067.
This yields $a^*_0 \sim$ 97.94 ${\rm \AA}$ and $H^*\sim$ 11.86 meV.

In cylindrical coordinates the KS equations read

\begin{eqnarray}
& & \left[-\frac{1}{2} \left( \frac{\partial^2}{\partial r^2}
+ \frac{1}{r} \frac{\partial}{\partial r} - \frac{l^2}{r^2}
+ \frac{\partial^2}{\partial z^2} \right)
+ V_{cf}(r,z) \right.
\nonumber
\\
& &
\label{eq1}
\\
&+& \left. V^H + V^{xc} + W^{xc}\,\eta_{\sigma} \right]
u_{n l \sigma}(r,z) =
\epsilon_{n l \sigma} u_{n l \sigma}(r,z) \,\, ,
\nonumber
\end{eqnarray}
where $\eta_{\sigma}$=$+1(-1)$ for $\sigma$=$\uparrow$$(\downarrow)$,
$V_{cf}(r,z)$ is the confining potential,
$V^H(r,z)$ is the direct Coulomb potential, and $V^{xc}={\partial
{\cal E}_{xc}(n,m)/\partial n}\vert_{gs}$ and
$W^{xc}={\partial
{\cal E}_{xc}(n,m)/\partial m}\vert_{gs}$
are the  variations of the exchange-correlation
energy density ${\cal E}_{xc}(n,m)$  in terms of the electron
density $n(r,z)$ and of the local spin magnetization
$m(r,z)\equiv n^{\uparrow}(r,z)-n^{\downarrow}(r,z)$ taken at the
ground state (gs).

As usual,
${\cal E}_{xc}(n,m) \equiv {\cal E}_{x}(n,m) + {\cal E}_{c}(n,m)$
has been built from 3D homogeneous electron gas
calculations. This yields a well-known\cite{Lun83}, simple analytical
expression for the exchange contribution ${\cal E}_{x}(n,m)$.
For the correlation contribution ${\cal E}_{c}(n,m)$ we have used
two different parameterizations, both based on the results of
Ceperley and Alder.\cite{Cep80} The first was proposed by
Vosko, Wilk and Nusair\cite{Vos80}, and the second by
Perdew and Zunger.\cite{Per81} We have checked that they yield
the same results, and all results presented in this
work have been obtained with the exchange-correlation energy density
proposed by Perdew and Zunger.

For a double QD the confining potential $V_{cf}(r,z)$
has been taken to be parabolic with frequency $\omega_0$
in the $xy$ plane, plus
a symmetric double quantum well in the $z$ direction.
For a single QD we have also used a parabolic confining potential
in the $xy$ plane, together with a
quantum well in the $z$ direction. Any
other axially symmetric $V_{cf}(r,z)$ can be implemented as well.

$V^H(r,z)$ was obtained solving the Poisson equation
using the conjugate gradient method\cite{Dav80,Pre92} (CGM). This
requires a knowledge of  $V^H(r,z)$ at the mesh boundary
which can be obtained  by direct integration.
Due to  axial symmetry
\begin{equation}
V^H(r,z) = 2 \int_0^{\infty}  r'\, dr'
\int_{-\infty}^{+\infty}   dz' \,\Delta n(\vec{r}\,')\,
[(r+r')^2+(z-z')^2]^{1/2} \,{\bf E}(\alpha^2) \,\,\, ,
\label{eq3}
\end{equation}
where {\bf E} is the complete elliptic integral of the second
kind\cite{Abr70} and $\alpha^2\equiv 4 r r'/[(r+r')^2 +(z-z')^2]$.

We have discretized   the KS and Poisson equations
using $k$-point Lagrange formulae for the $r$- and $z$-derivatives, and
$k+1$-point Lagrange formulae for the integrations.\cite{Abr70} The
mesh size has to be such that   the discretized wave
functions $u(r_i,z_j)$ at the mesh boundary can be safely taken as zero.
This is one boundary condition for  physically acceptable solutions
to Eq. (\ref{eq1}), the other one is the regularity of the sp
wave functions at $r = 0$. In our scheme the  $\Delta r$ and $\Delta z$
steps may have different values. The high precision demanded
by the calculation imposes restrictions on the possible values of $k$,
$\Delta r$, and $\Delta z$ and will be discussed below.
This space discretization scheme offers an efficient calculation of
sp wave functions, thus of the electron densities
and direct Coulomb and exchange-correlation potentials.

The  imaginary time method is described in detail in
Ref. \onlinecite{Dav80}. It is based on the observation that the
discretized time-evolution operator ${\cal H}(t)$ for
the time-dependent KS equations

\begin{equation}
\imath \hbar\, \frac{\partial \phi_j(t)}{\partial t} =
{\cal H}(t) \phi_j(t)
\label{eq5}
\end{equation}
which  formally yields

\begin{equation}
\Psi^{(n+1)}_j = \exp\left(-\frac{\imath}{\hbar} \Delta t {\cal H}^{(n)}
\right) \phi^{(n)}_j
\,\,\, ,
\label{eq6}
\end{equation}
where $\Delta t$ is the time step and $n$ indicates the step iteration,
for imaginary $\Delta t = -\imath \Delta \tau$
($\Delta \tau > 0$) produces a decrease of the KS energy.
In imaginary time, the wave functions $\{\Psi^{(n+1)}_j\}$ are no longer
orthonormal, and a Gram-Schmidt orthonormalization has to be carried
out after each  iteration to obtain $\{\phi^{(n+1)}_j\}$ from
$\{\Psi^{(n+1)}_j\}$. To first order in $\Delta \tau$, Eq. (\ref{eq6}) becomes

\begin{equation}
\Psi^{(n+1)}_j = \left(1- \frac{{\cal H}^{(n)}
\Delta \tau}{\hbar} \right) \phi^{(n)}_j \,\,\, ,
\label{eq7}
\end{equation}
which shows the simplicity of the method and its  practical
implementation: after discretizing the KS Hamiltonian and wave functions,
it  essentially reduces to  repeated application of the KS
Hamiltonian to the previous  wave functions.
Moreover, it is easy to check from Eq. (\ref{eq7}) that upon convergence

\begin{equation}
\epsilon_j^{(n+1)} = \frac{\hbar}{\Delta \tau} \left[
1 - \langle \phi_j^{(n)}|\Psi^{(n+1)}_j \rangle \right]
\label{eq8}
\end{equation}
coincide with the KS sp energies.
Equation (\ref{eq7})  shows that the ITM belongs
to the general class of relaxation methods employed to solve  partial
differential equations. This provides  a simple
criterion for  fixing
$\Delta \tau$ in such a way that the imaginary time evolution is
stable\cite{Pre92}, namely
$\hbar^2 \Delta \tau_{max} /(2 m \Delta^2) < 1/4$, with
$\Delta$  the smallest between the $r$ and $z$ steps.
In actual calculations we have taken
$\Delta \tau = 0.1 \,\Delta \tau_{max}$.

To start the ITM iteration  one needs a set of sp wave functions
and energies to build the initial sp level scheme. We have used two such
sets. The first consists of the wave functions of an axially
symmetric, 3D harmonic oscillator potential of frequencies $\omega_0$
and $\omega_z$ in the radial and $z$ directions, respectively. This
potential gives rise to  analytical solutions even in the presence of
a constant magnetic field in the $z$ direction, and is a suitable
 confining potential for a single QD. In this case, the sp
Hamiltonian is separable and the wave function can be written as $u(r,z) =
{\cal R}(r)\,{\cal Z}(z)/\sqrt{2\pi}$ with
\begin{eqnarray}
{\cal R}(r) &=& \frac{1}{a}
\sqrt{\frac{n_r!}{2^{|l|}(n_r+|l|)!}}
 \left(\frac{r}{{a}}\right)^{|l|}
e^{-(r/2a)^2} L^{|l|}_{n_r}\left(\frac{r^2}{2a^2} \right)
\nonumber
\\
& &
\label{eq9}
\\
{\cal Z}(z) &=& \sqrt{\frac{\zeta}{\sqrt{\pi}\,2^{n_z}\,n_z!}}\,
H_{n_z}(\zeta z) e^{-\zeta^2\,z^2/2}   \,\,\, ,
\nonumber
\end{eqnarray}
where $a \equiv \sqrt{\hbar/2m\omega_0}$,
$\zeta \equiv \sqrt{m\omega_z/\hbar}$, and $L^{|l|}_{n_r}$ and
$H_{n_z}$ are generalized Laguerre and Hermite
polynomials\cite{Abr70}, respectively. The sp energies are
${\cal E}_{n_r,n_z,l} = \hbar \omega_0 [ 2 n_r +|l| +1]+
 \hbar \omega_z [ n_z + 1/2]$
with $n_r$ and $n_z$ equal to 0, 1, \ldots

For a double QD, we have found  it convenient to choose
${\cal Z}(z)$ as the lowest energy eigenfunctions of a 1D double
quantum well.  The QD thickness is such that for not too many
electrons, only the two lowest states are needed.  If the double dot
is symmetric, these solutions are either even or odd under reflection
$z \rightarrow -z$. Actually, for one QD, ${\cal Z}(z)$ can also be
the wave function in the 1D quantum well.

Both for a single QD and for a symmetric double dot, the single-electron
wave functions
are characterized by the values of $l_z$, $s_z$ and parity, i.e.,
they must be either symmetric or antisymmetric under inversion
$\vec{r} \rightarrow -\vec{r}$, and  be either even or
odd when reflecting $z \rightarrow -z$. All these symmetries are
included
in the starting separable wave functions.
Indeed, the Hermite polynomials are even
or odd depending on $n_z$, and the parity of a sp level is simply
$(-)^{n_z+|l|}$. As the KS Hamiltonian and the ITM preserve these
symmetries,  in the course of the iteration
procedure the sp wave functions, which are no longer  separable,
keep their initial quantum numbers (orbital and spin projection on the
$z$ axis, parity and reflection symmetry with respect to the $z=0$ plane)
which are conserved quantities.

To check the numerical scheme we have carried out extensive and
systematic tests. A first test on the discretization and iteration
procedure  consisted in numerically solving
an axially deformed harmonic oscillator potential. We have exactly
reproduced the spectrum ${\cal E}_{n_r,n_z,l}$ given after
Eqs. (\ref{eq9}).
The implementation of the CGM has been successfully tested comparing
 results computed for a spherical Gaussian charge distribution with the
analytical distributions.

As another test of the numerical code, we have  compared
the total energy calculated from a straightforward integration of the
energy density
with the expression in terms of the sp energies derived
 from the KS equations. Writing the correlation
energy density\cite{Per81,Vos80}
as ${\cal E}_{c}[n,m] \equiv n\,{\cal E}_{cor}[n,\xi]$, where
${\cal E}_{cor}[n,\xi]$ is the correlation energy per electron and
$\xi \equiv m/n$ is the local spin polarization, one obtains

\begin{equation}
E = \sum_j \epsilon_j -
\,\int d\vec{r}\, \left\{ \,\frac{1}{2}\, V^H(\vec{r}\,)\,n(\vec{r}\,)
+\frac{1}{3}\,{\cal E}_{x}[n,m]
+\frac{\partial\,{\cal E}_{cor}[n,\xi]}{\partial n} \, n^2(\vec{r}\,)
 \right\}
\label{eq12}
\end{equation}
We have checked that in the worst case, the energies calculated with
either method
agree to within one part in $10^4$ for our combination of
$k=7$ formulae and the values $\Delta r= \Delta z = 0.12\, a^*_0$
used in this work.\cite{note1}
Simpler three point formulae ($k=3$) turned out to be inaccurate for
reasonable spatial steps, and   $k=9$ and 11 formulae, which
allow  larger  steps, did not  appreciably reduce the
 computing time.  It might be interesting to remark that most of
this disagreement  arises from the integral of the external
confining potential for a quantum well in the $z$ direction,
 due to the sharp discontinuity in the given potential. We have
checked that for a parabolic confining potential  in the
$z$ direction, both methods of evaluating the total energy
agree to within one part in $10^7$.

In the case of double QD's, we have
also checked that  the  results coincide irrespective of whether
we start the iteration from pure harmonic oscillator wave functions,
or from the better choice of  double quantum well eigenfunctions
in the $z$ direction. For even-$N$ systems we have always started the
iteration with non-identical sp potentials for spin-up and spin-down
electrons, to avoid artificial configurations with $S_z=0$.

\section{Results}

\subsection{Single quantum dot}

We have first addressed  the addition energies of
one quantum dot hosting up to $N=21$ electrons.
The confining potential in the $z$ direction is a
quantum well  $W=12$ nm
wide, which corresponds to the experimental well\cite{Aus98},
and $V_0=200$ meV deep.\cite{Ron98}
Electrons are laterally confined by the parabolic potential
$m \omega_0^2 r^2/2$ for which we have tried
different $\omega_0$ values.

Figure \ref{fig1} shows the addition spectrum
$\Delta A(N) \equiv E(N+1) -2 E(N) + E(N-1)$
for $\omega_0$ = 3, 5, and 10 meV. In addition to local maxima at
shell filling values $N$ = 2, 6, 12, and 20, other peaks appear at
half-filling values $N$ = 4 and 9. This is a consequence of
 Hund's rule, which establishes that degenerate
electronic states in a shell are filled with parallel
spins up to half-shell, in order to maximize the exchange
interaction.

Up to $N$ = 12 the results look very similar to
the experimental values\cite{Tar96}, especially for medium
and weak  $r$-confinement. For larger $N$
values we have found a conspicuous even-odd effect, instead
of the weak maxima experimentally found at $N$ = 20,
and especially at $N$ = 16.
This result does not mean that Hund's rule is violated
within this shell. Indeed, we have found spin alignment
up to $N$ = 16, but the associated energy gain is not
enough to produce the local maximum at $N$ = 16. It is
worth pointing out that the energies involved in the
definition of $\Delta A(N)$ are very large, compared with
the  second energy difference. For example, for
$\omega_0$
= 5 meV, $E(16)$ = 1.048 eV, whereas $\Delta A(16)$ is
$\sim$ 3 meV.

For strong $r$-confinement, our results are
quite similar to those in Ref. \onlinecite{Ron98}, except  for medium
and small values. This is not surprising, since in this Ref.
the 3D electron-electron energy is treated in first-order
perturbation theory\cite{Mer70}, thus  one should  not expect it to hold
 if the confinement is not strong enough.  The two-dimensional  $E(N+1) -
E(N)$ results at $B$ = 0 coincide with those in
 Ref. \onlinecite{Ste98} when $\omega_0$ = 5 meV.

Our 3D dot is rather strongly confined in the $z$
direction, and the results for $\Delta A(N)$ are similar to those
obtained for pure 2D dots.  We have  computed the addition energies
using the 2D dot model described in detail in Ref. \onlinecite{Pi98};
as  shown in Fig. \ref{fig2}, 2D and 3D results are very similar
and  qualitatively better than those from other
different 2D models. \cite{Mac97}
 As one increases the confinement in the
$z$ direction up to $V_0$ = 300 meV, the addition energies become
indistinguishable in the scale of  Fig. \ref{fig1}. We shall see
that this is not the case for double dots, and that even qualitative
differences appear if the dots are strongly coupled.

In spite of this quasi two-dimensional behaviour, the
electron density spreads beyond the well up to distances that
are relevant for vertically coupled dots. This is illustrated
in Fig. \ref{fig3}, where we have plotted the $n(z)$
electron density defined as

\begin{equation}
n(z) \equiv \int_0^{\infty} dr \, r \,n(r,z)
\label{eq14}
\end{equation}
corresponding to $N$= 12, $W$= 12 nm, and $V_0$ =
200 and 300 meV [notice that $N = 2 \pi \int dz \, n(z)$].
The electron densities spill out of the quantum well,
the smaller the confinement, the larger the effect.

Finally, in Fig. \ref{fig4} we represent the density
of a 2D dot hosting 12 electrons, and the $n(r)$
density of the corresponding 3D dot defined as
$n(r) = \int dz \, n(r,z)$
[notice that $N = 2 \pi \int dr \,r\,n(r)$], both
for $\omega_0$ = 5 meV. These radial densities
are very similar. The 3D densities $n(r)$ for
$V_0 = 300$
meV, instead of 200 meV, are indistinguishable within the
scale of the figure.

These comparisons allow one to infer
that the experimental dots are
quasi two-dimensional systems to a large extent, with moderate
lateral confinement, $\omega_0 \leq $ 5
meV.

\subsection{Double quantum dots}

We have modelled a symmetric double dot  by a parabolic
confining potential  with frequency
$\omega_0$ = 5 meV in the $r$
direction, and a symmetric double quantum well in
the $z$ direction. Each quantum well has a width equal to 12 nm,
and is separated from the other
by a barrier of thickness  $d$ that varies from 1 to 9 nm.
Some experimental results are available\cite{Aus98} for
this system at $d \geq$ 2.5 nm. Results for
well depths $V_0$ = 200 and 300 meV will be discussed.

Figure \ref{fig5} shows the $n(z)$ density profiles
corresponding to a quantum-mechanically coupled
configuration ($d$ = 2.5 nm), and to an electrostatically
coupled configuration ($d$ = 7.5 nm). Only in  the former case,
are the electrons delocalized.
A quantitative measure of this localization
is provided by the energy splitting between symmetric and
antisymmetric sp states (more precisely between even and
odd states with respect to specular reflection
$z \rightarrow -z$), $\Delta_{SAS}$. Indeed, for $N$ = 1 and
large distances, the symmetric and antisymmetric states
are degenerate and $\Delta_{SAS}$ approaches zero.
As $d$ decreases, the coupling increases and so does $\Delta_{SAS}$.
This effect depends weakly on $N$. We have plotted
$\Delta_{SAS}$  as a function of $d$ in Fig. \ref{fig6}, for the
lowest $l$ = 0 sp levels of systems with $N$ = 1 and 20.

Due to the characteristics of our double quantum well,
$\Delta_{SAS}$ depends exponentially on $d$,
$\Delta_{SAS}(d) = \Delta_0 \, \exp(-d/d_0)$.
For $V_0$ = 200 meV we have ($d_0, \Delta_0$) = (1.79 nm, 19.2  meV)
for $N$ = 1 and (1.76 nm, 17.6 meV) for $N$ = 20, whereas for $V_0$ =
300 meV, ($d_0, \Delta_0$) = (1.44 nm, 17.9  meV) for $N$ = 1 and
(1.42 nm, 16.5 meV) for $N$ = 20.  These values are similar to those
in Refs. \onlinecite{Aus98} and \onlinecite{Par00}.  It can be seen
from Fig. \ref{fig5} that for a given interdot distance, enlarging
$V_0$ decreases the coupling between the dots, as the density overlap
diminishes.  Figure \ref{fig7} shows the densities $n(r,z)$ for $N$ =
12 and for two $d$ values corresponding to quantal ($d$ = 2.5 nm) and
to electrostatic ($d$ = 7.5 nm) coupling.

Figure \ref{fig8} shows the addition spectrum for
quantum mechanically coupled ($d$ = 2.5 nm), and
electrostatically coupled ($d$ = 7.5 nm) dots, as well
as the results corresponding to the single 3D dot.
They have been obtained for $V_0$ = 200 meV.
Figure \ref{fig9} shows the same spectrum for
 $V_0$ = 300 meV. As expected, changes mostly
appear in the strong coupling case ($d$ = 2.5 nm).
Unexpectedly enough,  these changes are
qualitative, with, for example,  maxima in
$\Delta A(N)$ changed into minima.

Fig. \ref{fig6} indicates that at $d$ = 7.5 nm the
dots are well apart, only influenced by the
electrostatic coupling, and the addition energies
are insensitive to the value of $V_0$.
There are no experimental results in the literature for $\Delta A(N)$
at this interdot distance. However, the
experimental analysis\cite{Aus98} of the  derivative of the drain
intensity with respect to the drain voltage versus drain voltage
indicates that the electrons in the dots are indeed delocalized for
$d$ = 2.5 nm, and rather localized for $d$ = 7.5 nm.

The weak coupling at $d$ = 7.5 nm allows us to
interpret the appearance of several peaks in $\Delta A(N)$, such as
those at $N$ = 4, 8 and 12, as due to the fact that
2, 4, and 6 electrons on each dot already
yield maxima in the addition spectrum of a single QD.
The remaining peak at $N$ = 2
 corresponds now to half-filling the first shell
in each dot, which would close at $N$ = 2. This is
caused by the localization of one electron on
each constituent dot.\cite{Tok99}

Experimental results have been published\cite{Aus98,Tar98}
for $d$ = 2.5 nm. We have not attempted
to use $V_0$ as a fitting parameter,
so when comparing with experiment
one should bear in mind both the sensitivity of
$\Delta A(N)$ on the value of $V_0$ in the strong
coupling limit and  the results displayed in
Figs. \ref{fig8} and \ref{fig9}. Published calculations
correspond to depths lying in between these two
values.\cite{Ron99a,Par00}

One can see that maxima of $\Delta A(N)$ decrease on the
average when compared to the isolated QD. This is in
agreement with experiment. Notice however that
the experimental $\Delta A(N)$ for the strongly coupled
double dot\cite{Aus98} is  approximately half
 that corresponding to the single dot,
especially when $N \geq 10$. This feature is not reproduced
by the calculations.

Globally, the results for $d$ = 2.5 nm
are better reproduced with $V_0$ = 200 meV up to
$N$ = 12, and with $V_0$ = 300 meV for larger $N$ values.
Probably, a value of $V_0$ in between might improve the agreement
with experiment. Other possibilities such as
 an $N$-dependent $\omega_0$ might also be considered
\cite{Rei99,Par00}, or even some asymmetry in the double
well.\cite{Tok99}.
We have not tried these possibilities\cite{note2},
but we have checked that
self-interaction corrections (SIC), which are usually not
included in these kind of calculations, do not change the
addition spectrum. The results are presented in the Appendix.
%Consequently, the present results should be considered as
%a realistic model calculation to some extent.

For a given electron number, the gs configuration
may change as a function of the
barrier thickness. The new `phases', i.e. gs configurations
which appear as a function of $d$ have been thoroughly
discussed.\cite{Ron99a,Par00} To
label them, we have adopted the standard  convention of molecular
physics for sp electronic orbitals  as
$\sigma, \pi, \delta, \ldots,$ if $l= 0, 1, 2, \ldots$,
and upper case Greek letters are used for the total orbital angular
momentum. We have also used an adapted version\cite{Ron99a} of
ordinary spectroscopic notation
$^{2S+1}L_{g,u}^{\pm}$ where $S$ is the total
$|S_z|$, and $L$ is the  total $|L_z|$. The
superscript $+(-)$ refers to even (odd)
states under reflection with respect to the $z=0$ plane,
and the subscript $g(u)$ refers to positive (negative)
parity states.

We show  the evolution with  barrier
thickness of the energy and gs molecular configuration for
several $N$ values in Fig. \ref{fig10}.
The vertical lines have been drawn to guide the eye, and
different symbols have been used to identify different
phases. All panels in the figure display some common trends.
Initially, $E(d)$ increases with $d$. The reason is twofold.
On the one hand, at small $d$ all occupied sp levels
are specularly symmetric about the $z$ = 0 plane
$(`+$' states), the specularly antisymmetric levels
$(`-$' states) lie at much higher energies
(see Fig. \ref{fig6}).
On the other hand, the energies of the symmetric and antisymmetric state
respectively increase and decrease with $d$, and eventually both
states become degenerate at large interdot distances. This is a
well-known feature of the one particle, one-dimensional double quantum
well problem (see insert in Fig. \ref{fig6}), which remains valid in
the interacting many-electron calculation. The first phase transition
takes place when the first antisymmetric sp state becomes occupied.
At large distances, $E(d)$ slowly decreases with $d$ due to the
decrease of the interdot Coulomb energy. These trends are also
present in the Hubbard-like calculations of Ref.
\onlinecite{Ron99a}, but only the lowering of $E(d)$  due to the
interdot Coulomb energy is qualitatively  reproduced
by the calculations of Ref. \onlinecite{Par00}, which
fail to yield the energy growth at short distances.

In spite of the difference between the  values of the gs
energies reported in Refs. \onlinecite{Ron99a,Par00},
and also with respect to  the present work,
 which has to be mostly attributed to the different
confining potentials  in each
calculation, up to $N$ = 6 the phases are
the same but appear at different $d$ values. In this
respect, our results are in closer agreement with those
of Ref. \onlinecite{Par00}, possibly because
the radial frequencies $\omega_0$ are similar in both calculations.
We wish to point out that the phase diagram obtained
for $V_0$ = 300 meV is qualitatively similar to the one
shown in Fig. \ref{fig10}, but the phase transitions are
shifted $\sim $ 0.5 nm to the left.

\section{Summary}

We have used local spin-density functional theory to investigate
the zero magnetic field structure of one single and two
identical, vertically coupled QD's of finite
thickness.
While for one single dot, whose thickness corresponds
to that of  actual experimental devices, the addition
spectrum is quite similar to that predicted by
 purely two-dimensional models, in the case of
double dots, their vertical extension, is essential
for a quantitative description
of their quantum coupling, which influences the
addition spectrum at short distances.
For one single dot the calculated addition spectrum compares
well with experiment, whereas for two coupled dots the
agreement is  qualitative. This possibly reflects the fact
that in the latter case the spectrum is more sensitive to
the actual form of the bare confining potential.

The phase sequence of ground state configurations which
appear as a function of  interdot distance is
 quite similar to that found in previous
works\cite{Ron99a,Par00}, evolving from  the
`atomic phase' of two strongly coupled dots to the
`atomic phase' of two weakly coupled dots
through a series of `molecular-type phases' at intermediate
distances.
This is a rather robust picture, as it arises from the underlying
single-electron structure of the bare confining potential. Indeed,
the vertical confinement is so strong that at short distances
only symmetric sp states are occupied, the antisymmetric ones lying
at quite high energies. This  originates the atomic phase of
two strongly coupled dots.
 As the interdot distance increases the
$\Delta_{SAS}$ gap decreases, and the symmetric and antisymmetric
sp states eventually become degenerate, originating
the atomic phase of two weakly coupled dots.
The molecular-type configurations appear at intermediate distances
where $\Delta_{SAS}$ is similar to the other energy scale of the
system, namely $\hbar\,\omega_0$, and when the number of
electrons is large enough so that the system can minimize its energy
populating antisymmetric states.
The larger the number of  electrons, the larger the number
of populated antisymmetric states. This causes
the number of  intermediate molecular phases to
increase with $N$.

In spite of the mentioned qualitative agreement  with previous
results, the calculated phase diagrams are as
sensitive to the shape of the bare confining potential as the
addition spectra. It would be desirable that any prediction of the
actual appearance of
the phase diagrams should be based on a model that describes,
at least qualitatively, the corresponding
experimental addition spectrum.

\section{acknowledgments}

It is a pleasure to thank
Alfredo Poves, Flavio Toigo, and especially Guy Austing
for useful discussions.
This work has been performed under grants PB98-1247 and PB98-0124
of the DGESIC, and SGR98-00011 of the Generalitat of Catalunya. A.E.
acknowledges support from the Direcci\'on General de Ense\~nanza
Superior (Spain).

\appendix
\section{}

It is well known that the `exact' density functional for the
gs energy is self-interaction-free (see for instance
Ref. \onlinecite{Dre90}), but it is not the case of its current
approximations, such as LSDFT. One possible way of removing
this drawback is to use the self-interaction correction (SIC)
proposed by Perdew and Zunger\cite{Per81}, which introduces an
orbital-dependent single-particle potential, that
 improves  the total energy
of the electronic system and yields sp eigenvalues which approximate
the physical removal energies more closely, at the price of
rendering the KS minimization far more cumbersome. Since SIC's
are relatively more important for few electron systems, which
is the present case, we have tested their effect on the results shown
in the body of this paper, in two extreme configurations.
We refer the reader to Refs. \onlinecite{Per81} and
\onlinecite{Dre90} for a thorough
description of SIC. Here we only give the essential details of the
 application of the method to our physical problem.

Within the method of Perdew and Zunger,
the sp potential in Eq. (\ref{eq1}) becomes
orbital-dependent with the change

\begin{eqnarray}
V^H +  V^{xc} + W^{xc}\,\eta_{\sigma} &\longrightarrow &
\nonumber
\\
V_{eff} & \equiv &
 V^H + V^{xc} + W^{xc}\,\eta_{\sigma}
\nonumber
\\
 & - & V^H[n_{nl\sigma}] - V^{xc}[n_{nl\sigma},\, n_{nl\sigma}]
  - W^{xc}[n_{nl\sigma},\, n_{nl\sigma}] \,\eta_{\sigma}
\,\, ,
\label{eqa1}
\end{eqnarray}
where $n_{nl\sigma} = |u_{nl\sigma}(r,z)|^2$ is the
`orbital density' and
$V^H[n_{nl\sigma}]$ is obtained solving the Poisson equation
$\Delta V^H[n_{nl\sigma}] = - 4 \pi\, n_{nl\sigma}(r,z)$
as indicated in Section II. After selfconsistency is achieved, the
total energy can be obtained from Eq. (\ref{eq12}),
which has been written as

\begin{equation}
E \longrightarrow E -
\frac{1}{2}\,\sum_{nl\sigma}
\int d\vec{r}\; V^H[n_{nl\sigma}]\,n_{nl\sigma}(\vec{r}\,)
- \sum_{nl\sigma}
\int d\vec{r}\; {\cal E}_{xc}[n_{nl\sigma},\,n_{nl\sigma}]
\,\, .
\label{eqa3}
\end{equation}

We have drawn in Fig. \ref{figa1} the difference between
SIC and LSDFT addition energies defined as
$\mu(N) = E(N) - E(N-1)$
corresponding to the $V_0$ = 200 meV double quantum dot.
The average difference is small, of the
order of 0.4-0.5 meV. This difference
almost cancels out in the addition spectrum $\Delta A(N)$,
as  can be seen in Fig. \ref{figa2}. It
constitutes an interesting result in itself, indicating
that if one only wishes to  obtain the addition
spectrum, LSDFT does not need to be corrected for
self-interaction effects.

  \begin{figure}
   \caption[]{
     Addition energies $\Delta A$ as  functions of $N$ for a 3D
     single dot with $V_0$ = 200 meV, $W$ = 12 nm  and
     different $\omega_0$ values.
    }
   \label{fig1}
  \end{figure}

  \begin{figure}
   \caption[]{
Addition spectrum as a function of $N$ for a 2D
single dot and different $\omega_0$ values.
    }
   \label{fig2}
  \end{figure}

  \begin{figure}
   \caption[]{
$n(z)$ densities $[(a_0^*)^{-1}]$ for a 3D single dot with
$N$ = 12, width $W$ = 12 nm and well depths $V_0$ = 200 and 300 meV.
The vertical lines indicate the limits of the quantum well.
    }
   \label{fig3}
  \end{figure}

  \begin{figure}
   \caption[]{
$n(r)$ densities $[(a_0^*)^{-2}]$
for  2D and 3D single dots with
$N$ = 12 and $\omega_0$ = 5 meV. The  3D
well width and depth are  $W$ = 12 nm  and $V_0$ = 200 meV respectively.
   }
   \label{fig4}
  \end{figure}

  \begin{figure}
   \caption[]{
$n(z)$ densities $[(a_0^*)^{-1}]$ for an $N$ = 12 double dot of
width $W$ = 12 nm, well depth $V_0$, and
barrier thickness $d$. Top panel, $d$ = 7.5 nm. Bottom
panel, $d$ = 2.5 nm.
The vertical lines indicate the limits of the double quantum well.
    }
   \label{fig5}
  \end{figure}

\begin{figure}
\caption[]{
$\Delta_{SAS}(d)$ for the lowest  $l = 0$ sp
levels of a double dot  with $N$ = 1 and 20.
The insert shows the energies $E_S$ and $E_{AS}$ defining
$\Delta_{SAS}$, $\Delta_{SAS} \equiv E_{AS} - E_S$, for
one of the cases presented here.
}
\label{fig6}
\end{figure}

  \begin{figure}
   \caption[]{
$n(r,z)$ densities $[(a_0^*)^{-3}]$ for a double dot with
$N$ = 12, width $W$ = 12 nm, well depth $V_0$ = 200 meV, and
barrier thickness $d$. Top panel, $d$ = 2.5 nm. Bottom
panel, $d$ = 7.5 nm.
    }
   \label{fig7}
  \end{figure}

\begin{figure}
\caption[]{
Addition spectrum as a function of $N$ for a 3D
single dot (circles), and for two coupled dots
at $d$ = 2.5 nm (squares) and $d$ = 7.5 nm (diamonds).
The depth of the double quantum well is $V_0$ = 200 meV.
}
\label{fig8}
\end{figure}

  \begin{figure}
   \caption[]{
Same as Fig. \ref{fig8} for $V_0$ = 300 meV.
    }
   \label{fig9}
  \end{figure}

\begin{figure}
   \caption[]{
Energy and ground state molecular configurations
of the double dot as a function of the barrier thickness
for $N$ = 3 to 7.
The depth of the double quantum well is $V_0$ = 200 meV.
    }
   \label{fig10}
\end{figure}

\begin{figure}
   \caption[]{
Difference between  SIC and LSDFT addition energies
$\mu(N) = E(N)-E(N-1)$ for a double quantum dot of $V_0$ =200
meV depth and barrier thickness $d$ for $N =$ 1 to 13.
    }
   \label{figa1}
\end{figure}

\begin{figure}
   \caption[]{
Addition spectrum as a function of $N$
for two coupled dots
at $d$ = 2.5 nm (squares) and $d$ = 7.5 nm (diamonds).
The depth of the double quantum well is $V_0$ = 200 meV.
Solid symbols represent LSDFT results, and empty symbols
the results obtained after SIC's have been included.
    }
   \label{figa2}
\end{figure}

%\end{document}

\centering
\includegraphics[totalheight=197mm,width=139mm]{fig1.eps}
\centerline{Figure \ref{fig1}}
%   Figure 1
\centering
\includegraphics[totalheight=197mm,width=139mm]{fig2.eps}
\centerline{Figure \ref{fig2}}
%   Figure 2
\centering
\includegraphics[totalheight=197mm,width=139mm]{fig3.eps}
\centerline{Figure \ref{fig3}}
%   Figure 3
\centering
\includegraphics[totalheight=197mm,width=139mm]{fig4.eps}
\centerline{Figure \ref{fig4}}
%   Figure 4
\centering
\includegraphics[totalheight=197mm,width=139mm]{fig5.eps}
\centerline{Figure \ref{fig5}}
%   Figure 5
\centering
\includegraphics[totalheight=197mm,width=139mm]{fig6.eps}
\centerline{Figure \ref{fig6}}
%   Figure 6
\centering
\includegraphics[totalheight=197mm,width=139mm]{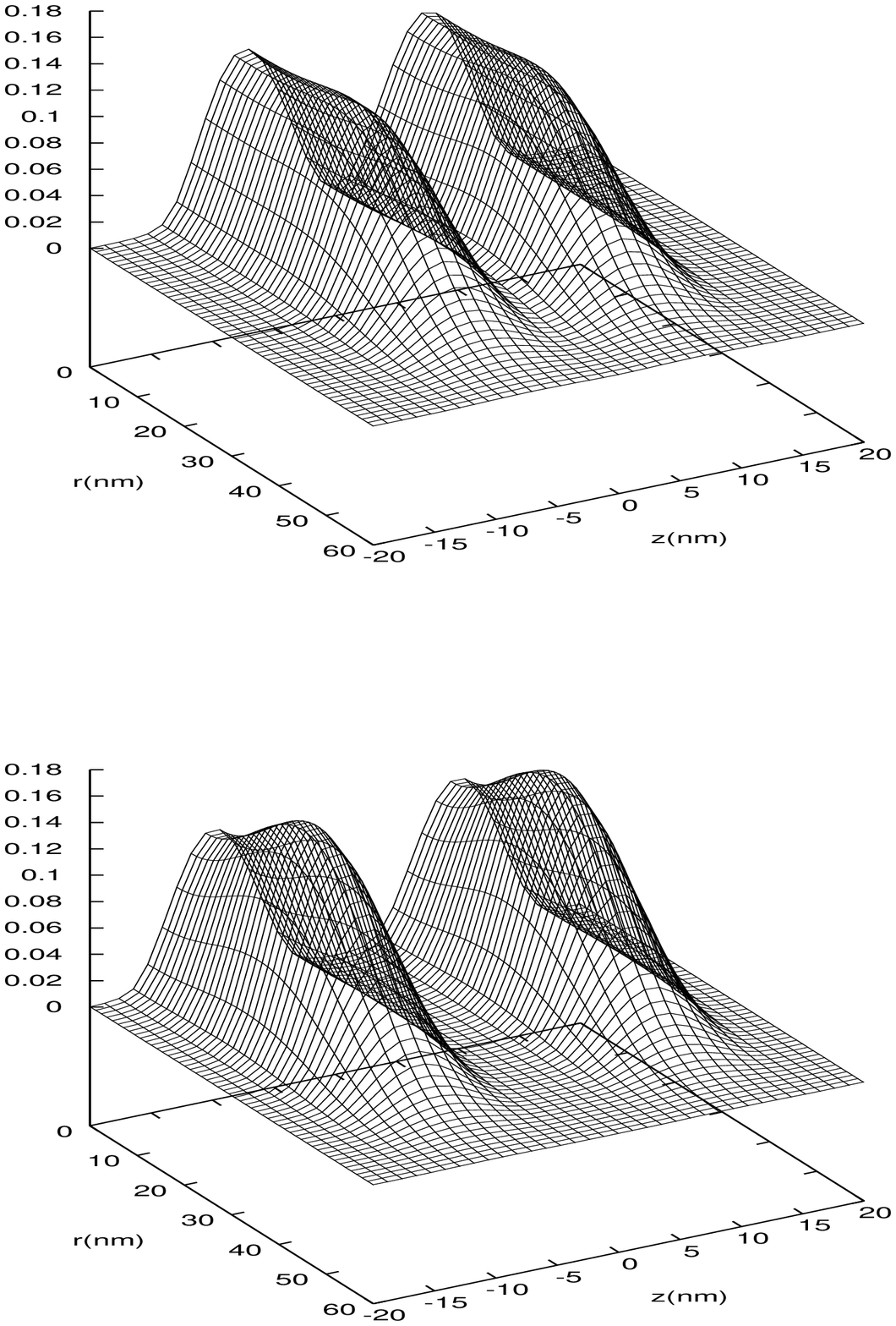}
\centerline{Figure \ref{fig7}}
%   Figure 7
\centering
\includegraphics[totalheight=197mm,width=139mm]{fig8.eps}
\centerline{Figure \ref{fig8}}
%   Figure 8
\centering
\includegraphics[totalheight=197mm,width=139mm]{fig9.eps}
\centerline{Figure \ref{fig9}}
%   Figure 9
\centering
\includegraphics[totalheight=197mm,width=139mm]{fig10.eps}
\centerline{Figure \ref{fig10}}
%   Figure 10
\centering
\includegraphics[totalheight=197mm,width=139mm]{fig11.eps}
\centerline{Figure \ref{figa1}}
%   Figure 11
\centering
\includegraphics[totalheight=197mm,width=139mm]{fig12.eps}
\centerline{Figure \ref{figa2}}
%   Figure 12

\begin{references}
\bibitem{note1} For the sake of comparison of, the $r$-step
we have routinely used\cite{Pi98} in 2D circularly symmetric
calculations is $0.06\, a^*_0$.
\end{references}
\end{document}